\documentclass{aa} 
\usepackage{graphicx}
\usepackage{txfonts}
\usepackage[T1]{fontenc}
\usepackage{siunitx}
\usepackage{hyperref}
\usepackage{xcolor}

\usepackage{subfigure}
\usepackage{soul}
\setstcolor{red}

\begin{document} 

   \title{The dynamical memory of tidal stellar streams}
   \subtitle{Joint inference of the Galactic potential and the progenitor of GD-1 with flow matching}

   \author{Giuseppe Viterbo
          \inst{1, 2},
          Tobias Buck
          \inst{1,2}
          }
   \institute{Universität Heidelberg, Interdisziplinäres Zentrum für Wissenschaftliches Rechnen (IWR), Im Neuenheimer Feld 205, 69120 Heidelberg,
Germany\\
 \and
 Universität Heidelberg, Zentrum für Astronomie, Institut für Theoretische Astrophysik, Albert-Ueberle-Straße 2, D-69120 Heidelberg, Germany\\
 \email{giuseppe.viterbo@iwr.uni-heidelberg.de}}

   \date{Received Month, 2/12/2025; accepted Month Day, 18/02/2026}

 \abstract
   {Stellar streams offer one of the most sensitive probes of the Milky Way’s gravitational potential, as their phase-space morphology encodes both the tidal field of the host galaxy and the internal structure of their progenitors. In this work, we introduce a framework that leverages flow matching and simulation-based inference (SBI) to jointly infer the parameters of the GD-1 progenitor and the global properties of the Milky Way potential.}
   {Our aim is to move beyond traditional techniques (e.g., orbit-fitting and action-angle methods) by constructing a fully Bayesian likelihood-free posterior over host galaxy parameters and progenitor properties, thereby capturing the intrinsic coupling between tidal stripping dynamics and the underlying potential.}
   {To achieve this, we generated a large suite of mock GD-1–like streams using our differentiable N-body code \textsc{\texttt{{Odisseo}}}, sampling self-consistent initial conditions from a Plummer sphere and evolving them in a flexible Milky Way potential model. We then applied conditional flow matching to learn the vector field that transports a base Gaussian distribution into the posterior $p(\theta \mid d)$, enabling efficient amortized inference directly from stream phase-space data.}
    {We demonstrate that our method successfully recovers the true parameters of a fiducial GD-1 simulation, producing well-calibrated posteriors and accurately reproducing parameter degeneracies arising from progenitor–host interactions. Our results highlight the power of modern generative models for dynamical inference and provide a scalable pathway toward jointly constraining Galactic structure and the origins of stellar streams.} 
   {Flow matching provides a powerful, flexible framework for Galactic archaeology. Our approach enables joint inference on progenitor and Galactic parameters, capturing complex dependencies that are difficult to model with classical likelihood-based methods. This work paves the way for fully simulation-driven dynamical inference using Gaia and upcoming surveys.}

\maketitle

\section{Introduction}
\label{sec:intro}
Assuming a hierarchical Lambda cold dark matter accretion history, the Galactic halo should be populated by tidal debris from accreted satellites, such as dwarf galaxies and star clusters \citep[e.g.,][]{Buck2019,Buck2020}. As described in \citep{biney_tremaine}, when the tidal forces acting on these objects are strong enough, stars are pulled out and end up on orbits that have slightly more or less energy than the progenitor's orbit, respectively forming the so-called leading and trailing arms. These narrow structures, referred to as stellar streams, have proven to be excellent tracers of fundamentals unknown in Galaxy evolution, and they have been used to probe the dark matter halo in external galaxies \citep{walder_probing_2024} and our Galaxy \citep{kupper_globular_2015}, chart dark matter subhalos in the Milky Way halo \citep{nibauer_textttstreamsculptor_2024}, and investigate the potential property of dark matter itself \citep{mestre_modelling_2024}. In particular, the long, dynamically cold GD-1 stream has been used to constrain the Milky Way potential with various techniques, including orbit fitting \citep{koposov_constraining_2010, malhan_constraining_2019}, backwards time integration \citep{price-whelan_inferring_2014, palau_constraining_2025}, 
action-angle modeling \citep{bovy_shape_2016}, action-angle clustering \citep{reino_galactic_2021}, and particle-spray with stream-track \citep{bowden_dipping_2015} modeling. More recently, new machine learning tools have been implemented to face this challenging problem, as in \citep{nibauer_charting_2022} and \citep{nibauer_galactic_2025}, where they estimated the acceleration felt by the stars in the stream as a probe of the Milky Way potential without relying on smooth analytic approximation. Notably, machine learning tools have opened a new avenue for more realistic representation of the potential. 

Recent developments in generative models have boosted the adoption of the simulation-based inference (SBI) technique \citep{Cranmer2020} as a valid alternative to the classical Bayesian approaches used in in the study of gravitational waves \citep{Dax_2025}, Galactic chemical enrichment \citep{buck_2025,Guenes2025}, Galactic archaeology \citep{Viterbo_2024, Sante_2025}, the dark matter density profile in dwarf galaxies \citep{Nguyen_2023}, and cosmology  \citep{Saoulis_2025}. In this work, we set out to jointly recover the gravitational potential of the Milky Way together with the parameters of the progenitor of the stellar stream in order to compare and extend the work presented in \citep{alvey_albatross_2024}.  
This paper is structured as follows. In Sect. \ref{sec:method} we present the simulation choices, the N-body simulator \texttt{\textsc{Odisseo}} used to create the training set, an introduction to the flow matching technique, and the details of the model architecture. In Sect. \ref{sec:results} we present the results of the inference for our reference GD-1 simulation and the test set results to validate the posterior calibration and accuracy. In Sect. \ref{sec:Conclusion} we summarize our findings and discuss limitations and future prospects. 
\begin{figure}
    \centering
    \includegraphics[width=\linewidth]{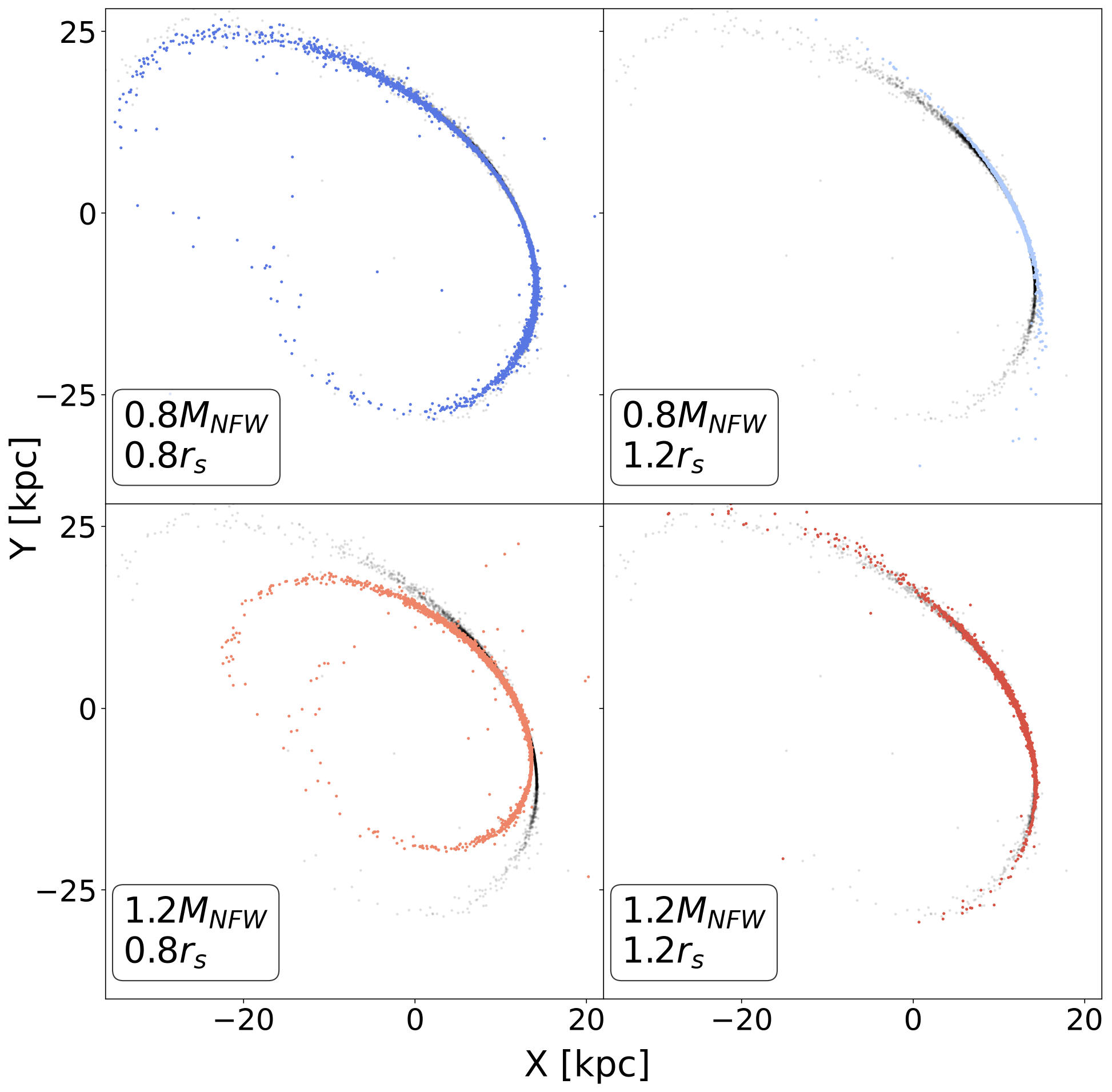}
    \caption{Stream morphology in different Galactic potentials. The black scatter plot is the output of the simulation for the fiducial values of \texttt{BovyMWPotential2014}, and the same is shown in each subpanel. We show the results for various combinations of 20\% offset of the mass of the NFW halo and its scale radius as indicated in the legend of each subpanel.}
    \label{fig:parameters_dependence}
\end{figure}

\section{Method}\label{sec:method}
Our goal is to recover, in a statistically sound way, the physical parameters, $\theta$, that govern the tidal stripping of a globular cluster evolving within a Milky Way–like gravitational field. Achieving this requires a set of modeling decisions regarding how we represent the host–galaxy potential (Sect. \ref{subsection:Galaxy}) and how we describe the progenitor system that seeds the stellar stream (Sect. \ref{subsection:Dwarf Galaxy}).
To infer $\theta$, we adopted a fully Bayesian framework. More specifically, we adopted a likelihood-free approach (SBI; Sect. \ref{subsec:sbi}). The SBI technique enables us to learn directly from simulations how the observed data, $d$, carry information about the underlying physical parameters. Specifically, we trained a generative model\footnote{A neural network designed to learn conditional probability distributions.} to approximate the posterior distribution, $p(\theta \mid d)$, by automatically extracting informative summary statistics from the observations.
The training dataset for this approach consists of pairs $(d^i, \theta^i)$, where each synthetic observation, $d^i$, is produced through forward modeling. We drew parameters, $\theta^i$, and passed them through a simulator, $S$ (\texttt{\textsc{Odisseo}}, Sect. \ref{subsection:odisseo}, yielding $d^i = S(\theta^i)$. The forward model must also incorporate observational caveats (e.g., uncertainty, selection effect) so that the forward process is equivalent to sampling from the likelihood $p(d\mid\theta)$.\footnote{For this reason, the SBI technique is also referred to as "implicit likelihood inference."} This simulation–inference amortized pipeline allowed us to connect theoretical models of tidal stripping with the observable properties of stellar streams in a principled and scalable way.

\begin{figure}
    \centering
    \includegraphics[width=1\linewidth]{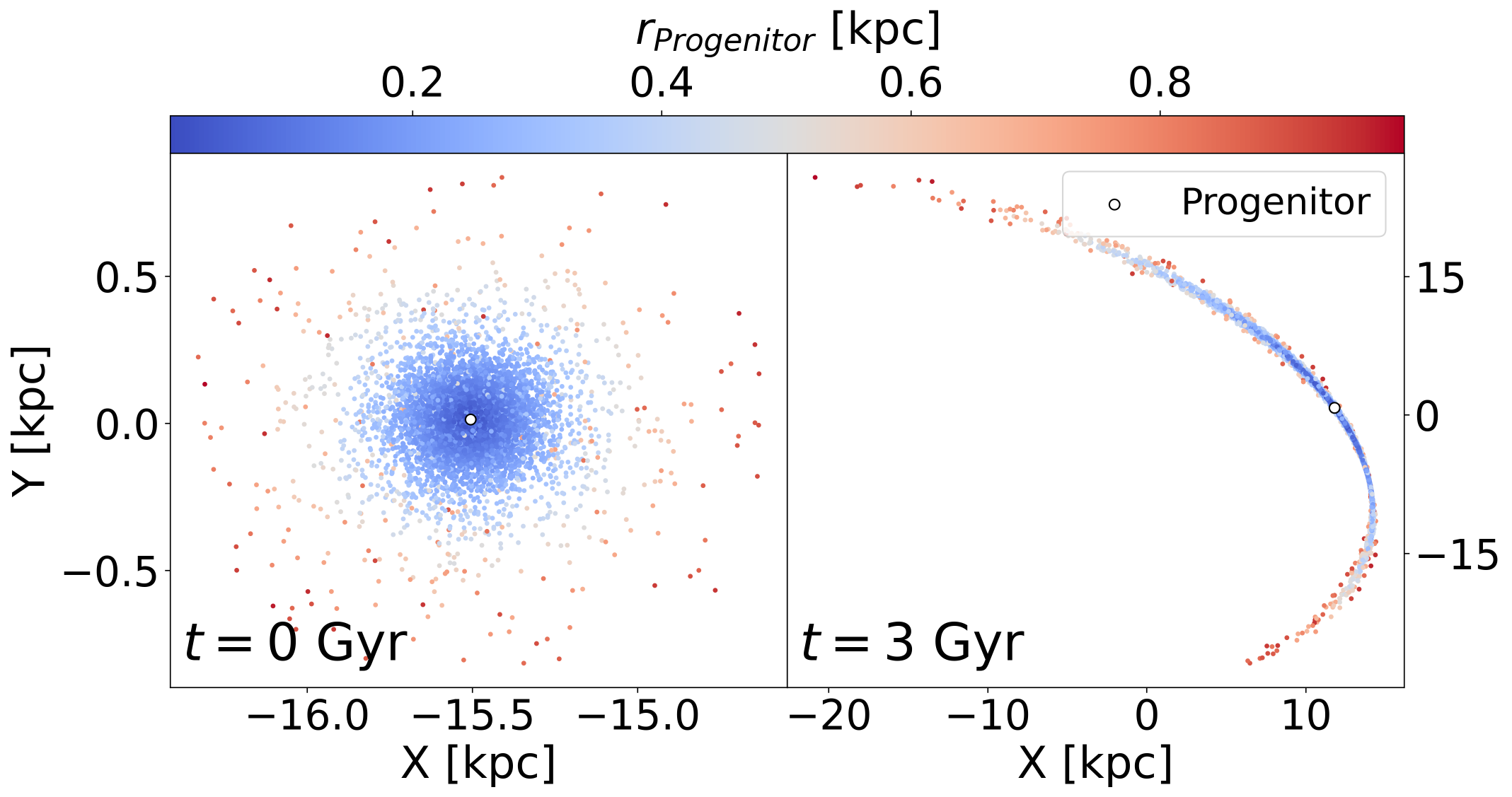}
    \caption{Tidal stripping of N=$10^4$ stars from a Plummer sphere over a 3 Gyr evolution. The parameters to set this simulation where the fiducial \texttt{BovyMWPotential2014} for the galaxy and $(M_{Plummer}, a_{Plummer}) = (10^{4.05} \text{M}_\odot, 100 \; \text{pc})$ for the progenitor. The color bar indicates the initial radial distance from the progenitor.}
    \label{fig:stripping}
\end{figure}

\subsection{The host: Model of the Milky Way}\label{subsection:Galaxy}
The gravitational potential of the Milky Way remains a subject of active debate. In the method outlined in the following section, we exploit the cold stellar stream GD-1 as a sensitive tracer of the Galaxy’s underlying density field. As illustrated in Fig. \ref{fig:parameters_dependence}, even when the properties of the progenitor are fixed, tidal stripping can give rise to strikingly different present-day stream morphologies, depending on the assumed Galactic potential. We adopted the widely used \texttt{BovyMWPotential2014} model \citep{bovy_dynamical_2014}, which provides a smooth and flexible yet tractable three-component representation of the Milky Way.

This framework consists of the following:
\begin{enumerate}
    \item A spherical dark-matter halo following a Navarro–Frenk–White (NFW) profile with density profile
    \begin{equation}
        \rho(r) = \frac{\rho_0}{\frac{r}{r_s} \left( 1 + \frac{r}{r_s} \right)^2}, 
    \end{equation}
    where $r$ is the spherical radial coordinate, $r_s$ is the scale radius, and $\rho_0$ is the central density.
    \item An axisymmetric disc described by a Miyamoto-Nagai (MN) potential 
    \begin{equation}
        \Phi(R,z)  = -\frac{GM_{MN}}{\sqrt{R^2 + (a+ \sqrt{z^2 + b^2})^2}},
    \end{equation}
    with $R, z$ being the radius and height in cylindrical coordinates, $M_{MN}$ as the total mass of the disc, and $a, b$ respectively as the scale length and height. 
    \item A spherical bulge with an exponential cutoff with a density profile of
    \begin{equation}
        \rho(r) = \rho_0 \left( \frac{r_1}{r_c} \right)^\alpha \exp{-(r/r_c)^2},
    \end{equation}
    with $r_c$ and $\alpha$ respectively as the cutoff radius and the power-law exponent.
\end{enumerate}

    \begin{figure}
        \centering
        \includegraphics[width=0.99\linewidth]{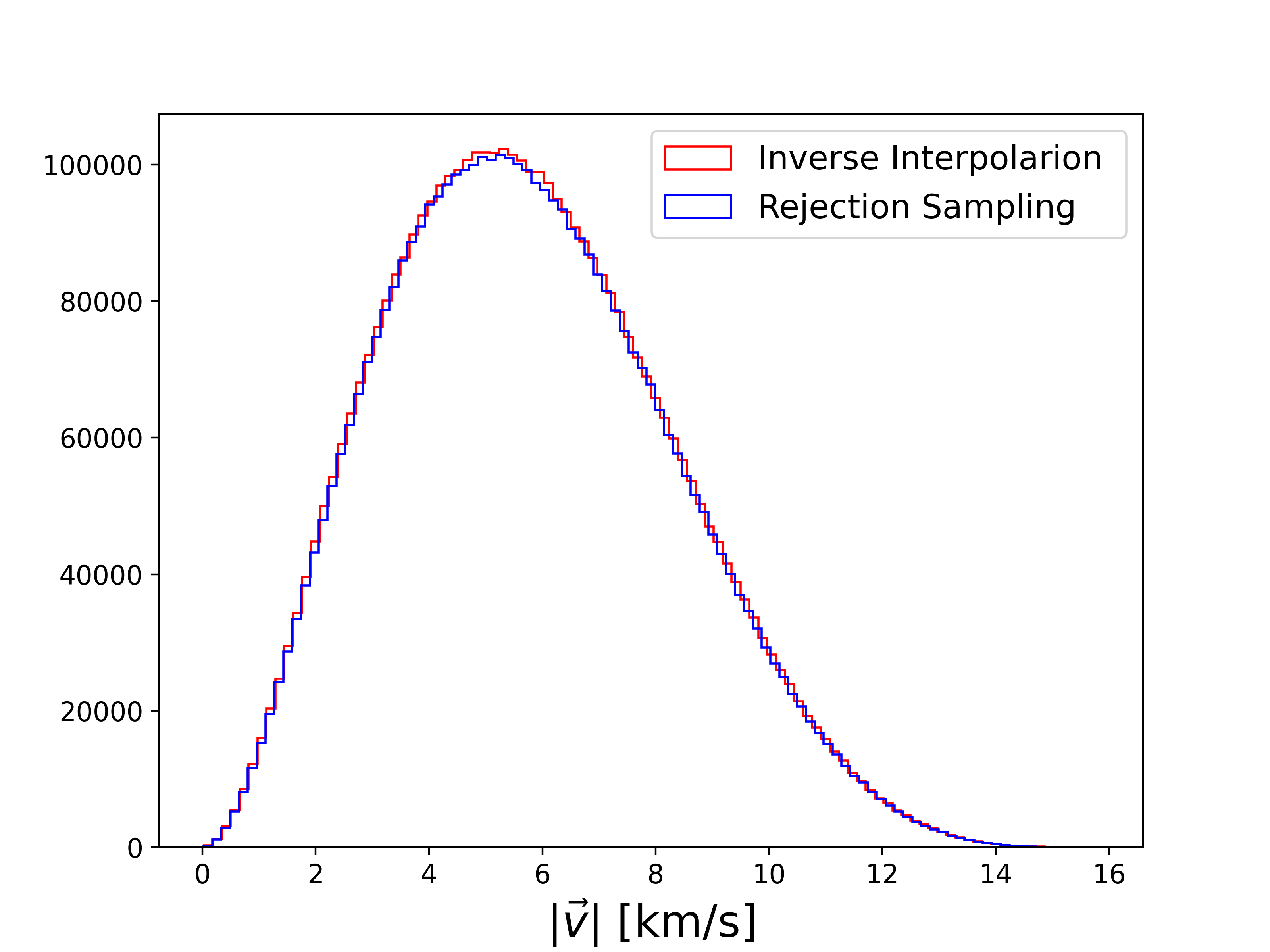}
        \caption{Comparison between rejection sampling and inverse sampling of the interpolated inverse for the module of the velocity of particles sampled from a Plummer sphere.}
        \label{fig:Rejection_vs_Inverse}
    \end{figure}
    
\subsection{The progenitor: Globular cluster}\label{subsection:Dwarf Galaxy}
The progenitor of GD-1 is widely attributed to be a disrupted globular cluster. Such systems are well approximated as spherically symmetric self-gravitating stellar ensembles \citep{aarseth_comparison_1974}, for which a Plummer sphere provides an analytically convenient and physically motivated model. The Plummer potential captures both the central concentration and the finite spatial extent expected for low-mass, pressure-supported clusters, making it ideally suited for generating initial conditions in stream-formation simulations.
A crucial aspect of modeling tidal disruption is that the internal phase-space distribution of stars within the progenitor directly shapes where and how stars are stripped. Stars located near the outer regions of the Plummer sphere—those with higher energies and larger orbital radii—reach the tidal boundary earlier and are therefore removed first as the cluster orbits within the Galactic potential. Conversely, tightly bound inner stars escape later and typically with different initial velocities \citep[see e.g.,][for a recent exploitation of this effect for Galactic Archaeology]{Skuladottir2025,Buder2025,Buder2025b}. 
We show an example of this process by evolving a Plummer sphere in the fiducial \texttt{BovyMWPotential2014} in Fig.\ref{fig:stripping}.

These variations imprint distinct escape conditions along the orbit, influencing the width, density variations, and the energy–angle structure of the resulting stream. Consequently, even subtle differences in the progenitor’s density profile or velocity dispersion propagate into observable differences in the stream morphology, making an accurate generative model of the progenitor essential for robust inference.
In the following section, we describe in detail how we sampled particles from a Plummer sphere to initialize the progenitor’s six-dimensional phase-space distribution. This sampling procedure forms one of the first steps of our forward model and ensures that the tidal stripping points—and thus the emergent stream structure—are physically consistent when generating the training data used for our SBI pipeline.

\paragraph{Differentiable initial condition.}\label{subsection:dif_initial_condition}
The initial positions and velocities of the progenitor depend inherently on the two parameters of the Plummer model: the total mass $M_{Plummer}$ and its scale radius $a_{Plummer}$.  To sample star particles from a Plummer sphere, we needed to sample ($x$, $v$) as follows:
\begin{enumerate}
    \item Given the radial mass profile of the Plummer model,
        \begin{equation}
            M(r) = M_{Plummer} \left(\frac{r}{a_{Plummer}} \right)^3 \left( 1+ \frac{r^2}{a_{Plummer}^2} \right)^{-3/2},
        \end{equation}
    we applied inverse sampling, using $F(r) = M(r)/M_{Plummer}$ as a cumulative distribution function to obtain $r$ by simply evaluating
        \begin{equation}
            r = \sqrt{\frac{a_{Plummer}^2}{u^{-2/3} -1}} \quad \text{where u$\sim \mathcal{U}(0,1)$}
        .\end{equation}
    \item Given the radial distance $r$ and assuming spherical symmetry, we generated ($x$, $y$, $z$) by sampling a random direction on the unit sphere
    \item Using the Plummer potential
        \begin{equation}
            \Phi_{Plummer} (r) = - \frac{GM_{Plummer}}{\sqrt{r^2 + a_{Plummer}^2}},
        \end{equation}
    we can find the associated escape velocity, $v_e(r)$, for each $r$ by simply setting the total energy of the particle to be 0, obtaining $v_e(r) = \sqrt{-2\Phi(r)}$. This is the maximum velocity that a gravitationally bound particle can have. In order to obtain the velocity, we used the distribution function, which for the Plummer sphere has the closed form
        \begin{equation}
            g(v)dv \propto (-E)^{7/2} v^2 dv,
        \end{equation}
    where $E =  -v_e^2 + \frac{1}{2}v^2$. We then defined $q(v) =  v/v_e$ so that the previous expression can be rewritten as the unnormalized probability density function:
        \begin{equation}
            g(q) = (1-q)^{7/2}q^2 \quad \text{with $0 \leq q \leq 1$}.
        \end{equation}
    In this case, inverse sampling is not possible because the cumulative distribution function $G(q) = \int_0^q (1-s)^{7/2}s^{2}ds$ does not have an analytic inverse, $F$, such that $F(G(q))=q$. However, since $G(q)$ can be evaluated numerically, we approximated its inverse, $\tilde{F}(u)$, with \(u \sim \mathcal{U}(0,1)\), by interpolating over a set of points \((G(q'), q')\) equally spaced values between zero and one. In practice, this amounts to swapping the input and output of \(G\) and constructing an interpolation of the inverse, which can then be sampled at the values of \(u\) to obtain our final samples for $q$. We report in Fig. \ref{fig:Rejection_vs_Inverse} a comparison plot of sampling $10^5$ particles, showing how close it matches the classical rejection sampling approach presented in \citet{aarseth_comparison_1974}. An advantage over the classical rejection sampling approach is that the inverse interpolation sampling is differentiable, allowing for gradient-based approaches, which we could leverage easily with our differentiable simulator, as shown in \citep{Viterbo_2025}. In this work however, we do not yet make use of the differentiability.   
    
    \item Once we had sampled $q$, we could rescale them using $v_e$ to obtain the velocity module $\vec{v}$. Finally, assuming spherical symmetry, we generated ($v_x, v_y, v_z$) by sampling a random direction on the unit sphere.

\end{enumerate}

\begin{figure*}
    \centering
    \includegraphics[trim=0cm 2.5cm 0cm 2cm, clip, width=1\linewidth]{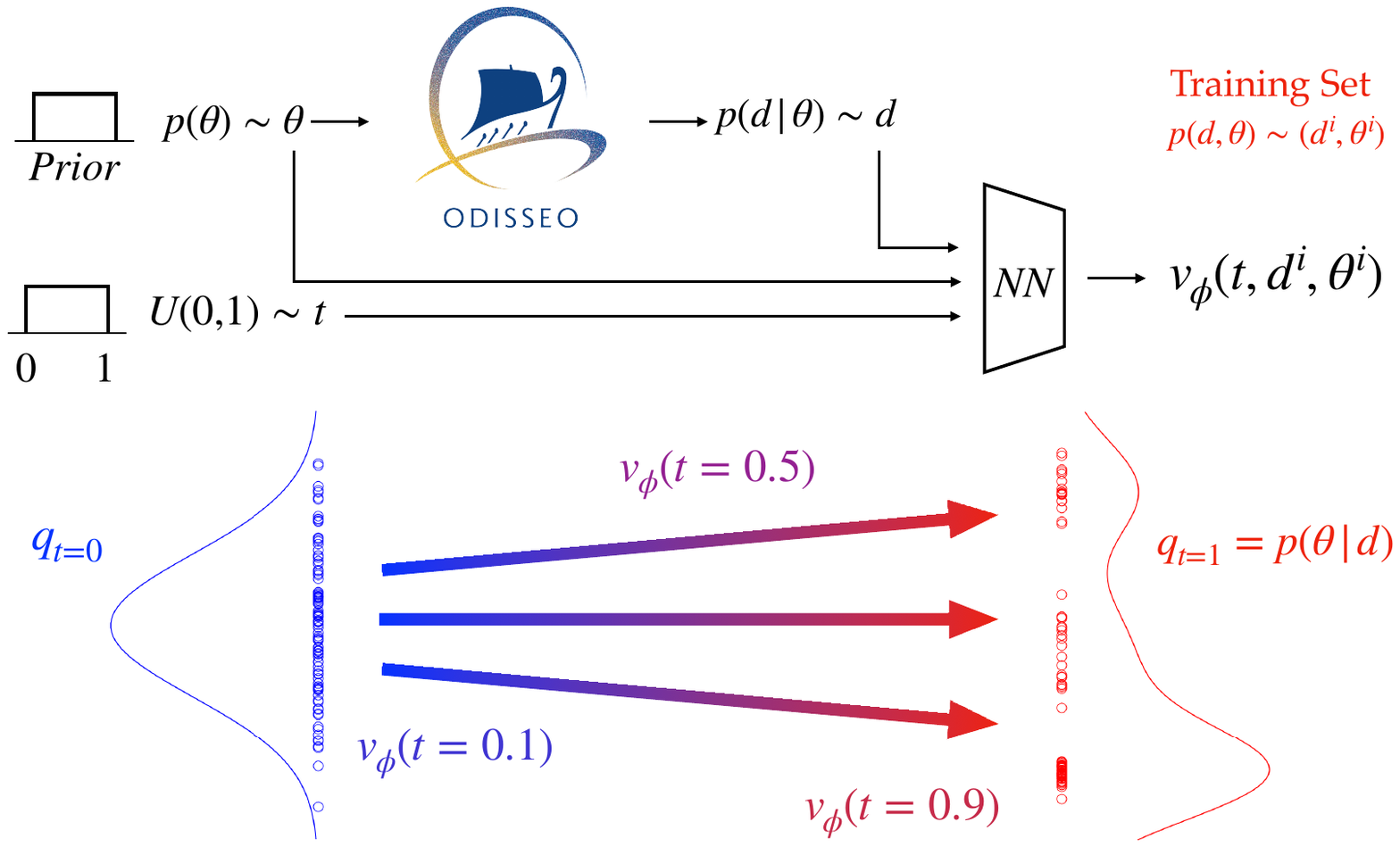}
    \caption{Schematic of flow matching for posterior estimation with \texttt{\textsc{Odisseo}}. The training set was generated by sampling  $i=0, ..., N$ parameters $\theta^i$ from the prior $p(\theta)$ and then forward modeled using \texttt{\textsc{Odisseo}} to obtain the observation $d^i \sim p(d\mid\theta)$. Following the flow described in Sect. \ref{subsection:flow_matching}, we sampled $t\sim U(0, 1)$ to train a neural network to approximate the vector field $v_\phi(t, d^i, \theta^i)$. In the lower section, we report a simplified flow matching objective for a 1D case. The neural network is called for different $t$ to regress the vector field that governs the ODE in order to transform the sampling distribution $q_{t=0} = \mathcal{N}(0, I)$ into the posterior distribution $q_{t=1} = p(\theta \mid d)$. Note that in this schematic, we refer to $\theta_1$ described in Sect. \ref{subsection:flow_matching} as $\theta$.}
    \label{fig:flow_matching_flowchart}
\end{figure*}

\subsection{N-body simulation with \texttt{\textsc{Odisseo}}}\label{subsection:odisseo}

Although fast particle-spray algorithms exist (e.g., \citealt{Chen_2025}; \citealt{Fardal_2015}; \citealt{nibauer_textttstreamsculptor_2024}), we modeled tidal disruption directly using the direct \texttt{\textsc{Odisseo}} N-body integrator. This self-consistent approach captures the ejection dynamics due to close encounters that particle-spray techniques approximate,\footnote{To keep the computational cost tractable, we nevertheless adopted a small Plummer softening length, as described in the following section.} and it produces forward simulations well suited for a simulation-based inference pipeline. As presented in \citep{Viterbo_2025}, \texttt{\textsc{Odisseo}} is a N-body code developed to study particle systems in which external potentials play a crucial role. The code is implemented in a purely functional style in \texttt{Jax} \citep{jax2018github}, which ensures that the full simulation is trivially parallelizable on GPU, and the high-level Python interface enables easy setup, prototyping, community-driven development, and maintenance. In addition, \texttt{\textsc{Odisseo}} natively supports just-in-time compilation and execution on CPUs, GPUs, and TPUs, ensuring both flexibility and computational efficiency. We used \texttt{\textsc{Odisseo}} as our simulator to quickly generate the training set of mock GD-1 simulations that are needed to train the model used in our SBI approach. Additionally, as described in \citep{Viterbo_2025}, the simulation can be differentiated by automatic differentiability (AD), allowing for gradient descent methods or variational inference for parameter estimation. In particular, the gradient can be pulled through the time integration and through the initial condition sampling, as described in \ref{subsection:Dwarf Galaxy}. In this work, we do not use the differentiability.

\begin{table}[t]
\caption{Prior ranges and true values for model parameters.} 
\centering
\begin{tabular}{lll}
\hline
\textbf{Parameter} & \textbf{Prior range} & \textbf{True value} \\
\hline
    & \textbf{Host Parameters} &    \\
\hline
\\
$M_{\mathrm{vir}}$ [$M_\odot$] & $[\,0.5\,M_{\mathrm{vir}}^{\mathrm{true}},\;2.0\,M_{\mathrm{NFW}}^{\mathrm{true}}\,]$ & $4.37\times 10^{11}$ \\
$r_{NFW}$ [kpc] & $[\,0.5\,r_{s}^{\mathrm{true}},\;2.0\,r_{s}^{\mathrm{true}}\,]$ & $16.0$ \\
$M_{\mathrm{MN}}$ [$M_\odot$] & $[\,0.5\,M_{\mathrm{MN}}^{\mathrm{true}},\;2.0\,M_{\mathrm{MN}}^{\mathrm{true}}\,]$ & $6.82\times 10^{10}$ \\
$a_{\mathrm{MN}}$ [kpc] & $[\,0.5\,a_{\mathrm{MN}}^{\mathrm{true}},\;2.0\,a_{\mathrm{MN}}^{\mathrm{true}}\,]$ & $3.0$ \\
\\
\hline
    &   \textbf{Progenitor Parameters}  &   \\
\hline
\\
$t_{\mathrm{end}}$ [Gyr] & $[\,0.5,\;5.0\,]$ & $3.0$ \\
$M_{\mathrm{Plummer}}$ [$M_\odot$] & $[\,10^{3.0},\;10^{4.5}\,]$ & $1.12\times 10^{4}$ \\
$a_{\mathrm{Plummer}}$ [kpc] & $[\,0.5\,a_{\mathrm{Plummer}}^{\mathrm{true}},\;2.0\,a_{\mathrm{Plummer}}^{\mathrm{true}}\,]$ & $0.008$ \\
$x^c$ [kpc] & $[\,10.0,\;14.0\,]$ & $11.8$ \\
$y^c$ [kpc] & $[\,0.1,\;2.5\,]$ & $0.79$ \\
$z^c$ [kpc] & $[\,6.0,\;8.0\,]$ & $6.4$ \\
$v_{x}^c$ [km s$^{-1}$] & $[\,90.0,\;115.0\,]$ & $109.5$ \\
$v_{y}^c$ [km s$^{-1}$] & $[\,-280.0,\;-230.0\,]$ & $-254.5$ \\
$v_{z}^c$ [km s$^{-1}$] & $[\,-120.0,\;-80.0\,]$ & $-90.3$ \\
\\
\hline
\end{tabular}
\tablefoot{The prior ranges for $t_{end}, M_{Plummer}, \vec{x}_c^0, \vec{v}_c^0$ are taken from \citep{alvey_albatross_2024}.}
\label{tab:priors}
\end{table}

\paragraph{Stellar stream simulations.}\label{subsection:simulation_steps}
In this section we describe the general steps that were used to generate simulations of a stellar stream. The steps are as follows:
\begin{enumerate}
    \item \textit{Cluster trajectory}. Given the present-day position and velocity $(\vec{x}^c, \vec{v}^c)$ for the progenitor of the star cluster, we traced the trajectory back to $t_{end}$ as a single particle with mass $M_{Plummer}$ in the chosen gravitational potential to the initial phase space position $(\vec{x}^c_0, \vec{v}^c_0)$.  
    \item \textit{Populate with star particles}. We drew $N$=1000 star particle positions and velocities $(\hat{x}_i, \hat{v}_i)_{i=0, \dots, N}$ from a Plummer potential centered on the origin, and then we shifted them in phase space by $(\vec{x}^c_0, \vec{v}^c_0)$. 
    \item \textit{Stream evolution}: We evolved the star particles forward in time for a total integration time of $t_{end}$ using the fifth order explicit Runge--Kutta method \texttt{Tsit5} ordinary differential equation (ODE) solver available in \texttt{diffrax} \citep{kidger_neural_2022}.
\end{enumerate}

All the simulations were carried out using \texttt{\textsc{Odisseo}}.\footnote{The data are publicly available at \url{https://zenodo.org/records/17711491}.} We decided to treat the particles as phase-space tracers, so we used a Plummer softening of 0.1 pc to avoid the formation of dynamical binaries.  

\subsection{Inference}\label{subsec:sbi}
Our aim was to jointly infer the parameters that describe the progenitor $(\theta_{prog})$ of a fiducial GD-1 stream simulation and the parameters of the Milky Way potential $(\theta_{host})$ in which the tidal stripping of its progenitor has happened. In practice, we aimed to both reproduce the results obtained by \cite{alvey_albatross_2024} and to extend the inference to also be able to use the stellar stream as a tracer for the gravitational potential of the host galaxy. We decided to face this challenging task by training a neural density estimator (\citealt{Cranmer2020}) of the posterior distribution $p(\theta \mid d_{obs})$, where $\theta = (\theta_{prog}, \theta_{host})$ and $d_{obs} = (\phi_1, \phi_2, r, v_{\phi_1}\cos{\phi_{2}}, v_{\phi_2}, v_{r})$ is the phase space of all the stars in the GD-1 stellar stream projected on the plane of the stream, as in \cite{alvey_albatross_2024} and \cite{koposov_constraining_2010}.\footnote{These are the standard set of co-ordinates used in the literature: ($ \phi_1, \phi_2$) are two angles coordinate, the corresponding proper motion $(v_{\phi_1}, v_{\phi_{2}})$, and radial distance and velocities $(r, v_r)$.} Since we modeled the progenitor of the GD-1 with a Plummer sphere (Sect. \ref{subsection:Dwarf Galaxy}), the parameters are the total time of integration, total mass, and the scale radius of the Plummer sphere, its present day position and velocity so that $\theta_{prog} = (t_{end}, M_{Plummer}, a_{Plummer}, \vec{x}^c, \vec{v}^c)$. Since we wanted to test the constraining power over the amplitude and shape of the host potential, we only varied the NFW halo mass and scale radius and the MN mass and scale length while fixing all the other parameters to the default fiducial parameters in \texttt{Galax} \citep{nathaniel_starkman_galacticdynamicsgalax_2024} for \texttt{MWPotential2014} in order to contain the computational cost of sampling on a small but still informative parameter space, leaving for future work extension to models with higher degrees of freedom, such as triaxial NFW profiles or time-evolving potentials. The final host potential parameters were then $\theta_{host} = (M_{NFW}, r_{NFW}, M_{MN}, a_{MN})$. 
To train a neural density estimator we needed to generate pairs of parameters and an observation $(\theta, d) \sim p(\theta, d)$ by letting $\theta \sim p(\theta)$, with $p(\theta)$ being the prior over the parameters, and $d = S(\theta) \sim p(d\mid\theta)$, with $S$ being the \texttt{\textsc{Odisseo}} simulator that implicitly defines the likelihood $p(d\mid\theta)$. Moreover, we applied to the observation the same observational window and noise level reported in Table 2 in \citep{alvey_albatross_2024}, leaving the background contamination and selection functions effect for future work. The prior choice is reported in Table \ref{tab:priors}.

\begin{figure}
    \centering
    \includegraphics[width=0.99\linewidth, trim=0.5cm 2.5cm 13.5cm 2cm]{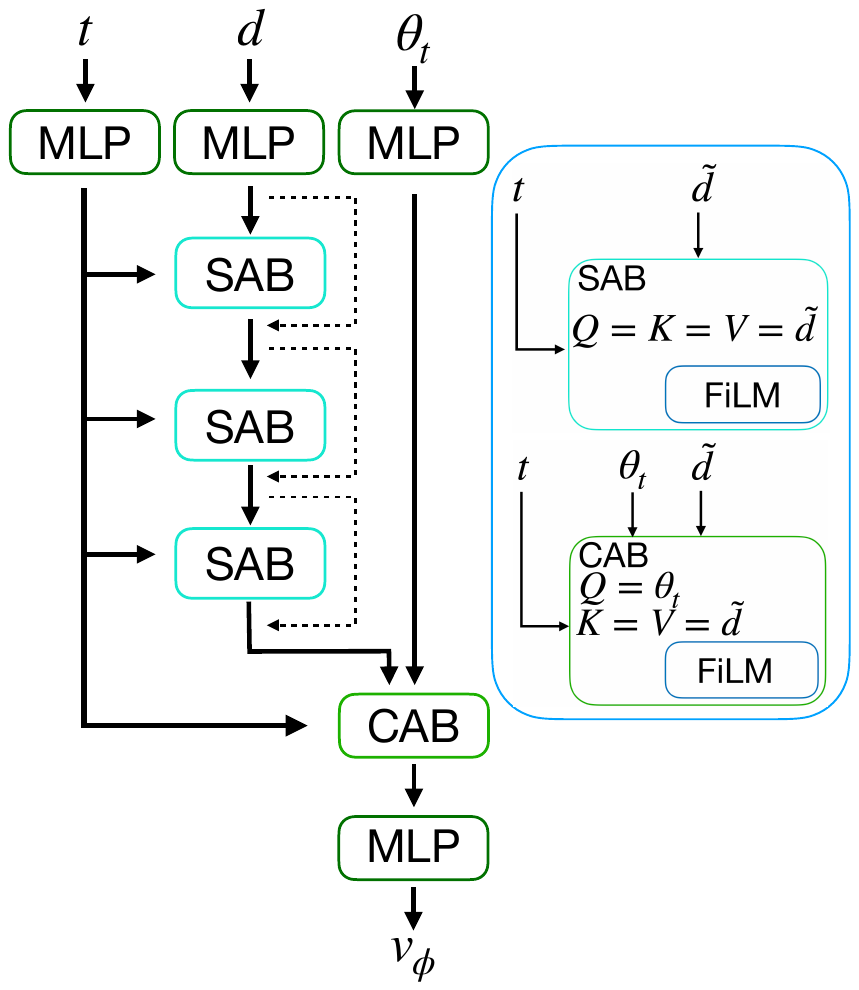}
    \caption{Flow matching \texttt{SetTransformer}. We indicate with $\tilde{d}$ both the observation ($d$) and the intermediate output of the attention blocks. To embed each of the stars in the observation ($d$) the parameter state $\theta_t$ at the ODE time $t$, we used a simple MLP with 128 neurons with a \texttt{SiLU} activation function. We then passed $\tilde{d}$ through three stacked SABs with a skip connection (dashed line) to encode the correlations between the particles while modulating the output of each block on $t$ using FiLM. Then we used a CAB with a FiLM modulation to focus the attention mechanism on finding the relevant feature in $\tilde{d}$ to regress the parameters $\theta$. The vector field $v_\phi$ was obtained by compressing the output of the CAB through an MLP with output dimensions equal to the dimensionality of $\theta$, in our case 13 dimensions.}
    \label{fig:SetTransformer}
\end{figure}

\paragraph{Flow matching posterior estimation.}\label{subsection:flow_matching}

\begin{figure*}[t]
    \centering
    \includegraphics[width=0.99\linewidth]{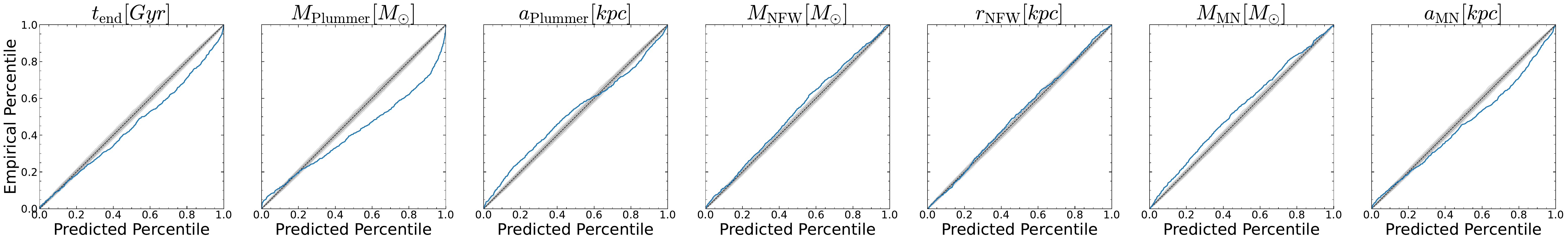}
    \vspace{0.5em} 
    \includegraphics[width=0.99\linewidth]{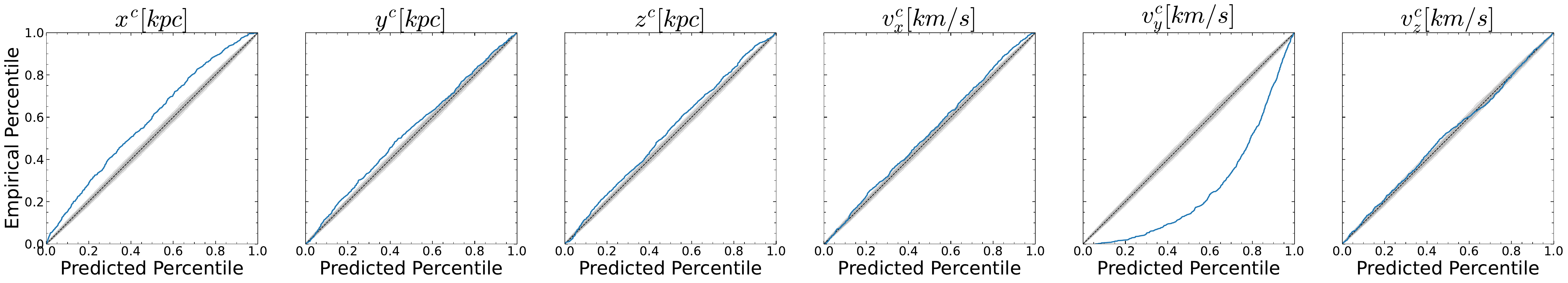}
    \caption{Percentile-percentile plot for the marginal posterior distributions over the test set. }
    \label{fig:PP_plot}
\end{figure*}

In recent years, many SBI applications, performed using normalizing flow architecture as a neural posterior estimator, have been proposed in the field \citep{Nguyen_2023,Sante_2025,Viterbo_2024,ltuili_2024}. The core idea is to leverage the flexibility of a neural network to learn a series of invertible transformations, conditioned on the observations, that project the parameters in a latent space where it is easy to sample from. In this way, instead of relying on the expensive "gold standard" Mark chain Monte Carlo technique, the transformations, via the change of variable formula, take care of tracking how the parameter space has been changed through the flow to the sample and evaluate the posterior distribution. The expressivity of this technique is limited by the necessity of using an invertible transformation \citep[usually spline functions as described in][]{durkan_neural_2019}. For this reason, neural posterior score estimation (\citealt{Geffner2022}) and flow matching posterior neural estimation (FMPE; \citealt{dax_flow_2023}) tackle the problem of approximating a conditional distribution with a different approach. As described in \cite{dax_flow_2023}, inspired by the promising results in generative tasks for which they were developed, these models transform noise into samples via trajectories parametrized by a continuous time variable $t$. In particular, for FMPE, the goal is to regress the vector field $v_\phi$, parametrized by a neural network with weights $\phi$, that describes these trajectories by solving an ODE. The key advantage over normalizing flows is that by regressing on the vector field $v_t$, we are free in the choice of the network architecture, at the cost of multiple network passages for sampling, since we need to solve the ODE for $t \in [0, 1]$. Moreover, contrary to neural posterior score estimation, which needs to solve a Stochastic differential equation, we are also able to track the posterior density directly.
As described in Sect. \ref{subsec:sbi}, we respectively trained observations and parameters on tuples ($d$, $\theta$). Our objective was to learn a continuous transformation for $\theta$ (or equivalently a probability path $q_t$) with constraints $\theta \sim q_0$ and $\theta \sim q_1=p(\theta \mid d)$, where $q_0$ is a sampling distribution, in our case a normal distribution with zero mean and identity covariance matrix. The ODE that controls this continuous process can be expressed as
\begin{equation}
    d\theta_t = v_{\phi}(t, d, \theta_t)dt.
\label{eq:ODE}
\end{equation}
Flow matching actually falls in the paradigm of continuous normalizing flow, but with an alternative objective function. In fact, since continuous normalizing flows are trained using a negative log likelihood as objective, one must track at training time the computationally expensive $p(\theta\mid x)$ by
\begin{equation}
    p(\theta \mid x)= q_1 = q_0 \exp{\left( \int_0^1 \nabla \cdot v_{\phi}(t, d, \theta)dt \right)},
\end{equation}
making the training of these models infeasible. As described in \citep{dax_flow_2023}, the key of flow matching is to directly regress the vector field $v_\phi$ on a vector field $u_t$ that generates the desired probability path $p_t$ while avoiding the integration of the ODE at training time.\footnote{The integration is still needed at inference time.} The nontrivial solution on how to perform this task, presented in \cite{lipman_flow_2023}, is to choose this path depending on the sample $\theta_1$\footnote{We indicate with $\theta_1$ the samples $\theta$ to be consistent with the literature on flow matching.} for which we want to model the probability path. In practice, we modeled the conditioned probability path $q_t(\theta \mid \theta_1)$ and the corresponding vector field $u_t(\theta \mid \theta_1)$ with the sample conditional flow matching (CFM) loss 
\begin{equation}
    \mathcal{L}_{CFM} = \mathbb{E}_{t\sim U[0, 1], d\sim p(d \mid \theta), \theta_t \sim q_t(\theta_t \mid \theta_1)} \| v_\phi(t, d, \theta_t) - u_t(\theta_t \mid \theta_1) \|^2.
\end{equation}
One simple choice for the conditioned probability path is the Gaussian path family
\begin{equation}
    q_t(\theta \mid \theta_1) = \mathcal{N}(\theta \mid t\theta_1, (1-(1-\sigma_{min})t\mathbb{I})
,\end{equation}
which generates the velocity field
\begin{equation}
    u_t(\theta \mid \theta_1) = \frac{\theta_1 - (1-\sigma_{min})\theta}{1-(1-\sigma_{min})t}, 
\end{equation}
with $\sigma_{min}>0$. These choices lead to this problem coinciding with the optimal transport \citep{lipman_flow_2023} between two Gaussian distributions: the sampling distribution and linear trajectory $t\theta_1$, ending in $\theta_1$ and with a smoothing constant given by $\sigma_1$. The steps to be followed to train boil down to sampling $\theta_1 \sim p(\theta)$ and transporting a point $\theta_0 \sim \mathcal{N}(0,\mathbb{I})$ from the sampling distribution to the posterior distribution on the linear trajectory $t\theta_1$ and ending in $\theta_1$. 

In the upper part of Fig.~\ref{fig:flow_matching_flowchart}, we summarize the SBI task. We first sampled from the prior parameters $\theta$, then forward modeled them using \texttt{\textsc{Odisseo}} to get the observation $d$, and used the couples $(d, \theta)$ as our training set. In the lower part of Fig.~\ref{fig:flow_matching_flowchart}, we show a representation of what the inference task looks like. We modeled the vector field $v_\phi$ with a neural network that is called for each integration step ($t \in [0, 1]$) of the ODE that transport the sampling distribution $\mathcal{N}(0, \mathbb{I)}$ into the posterior distribution $p(\theta \mid d)$.

\begin{figure}
    \centering
    \includegraphics[width=0.80\linewidth]{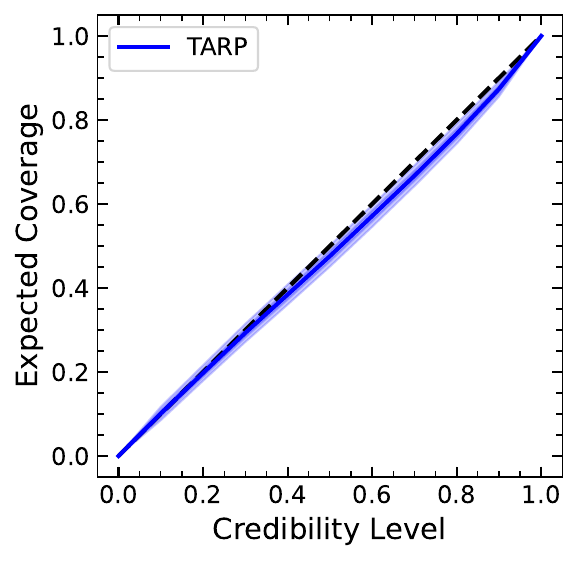}
    \caption{Tarp plot for the joint posterior distribution over the test set.}
    \label{fig:tarp}
\end{figure}

\begin{figure*}[t]
    \centering
    \includegraphics[width=0.99\linewidth]{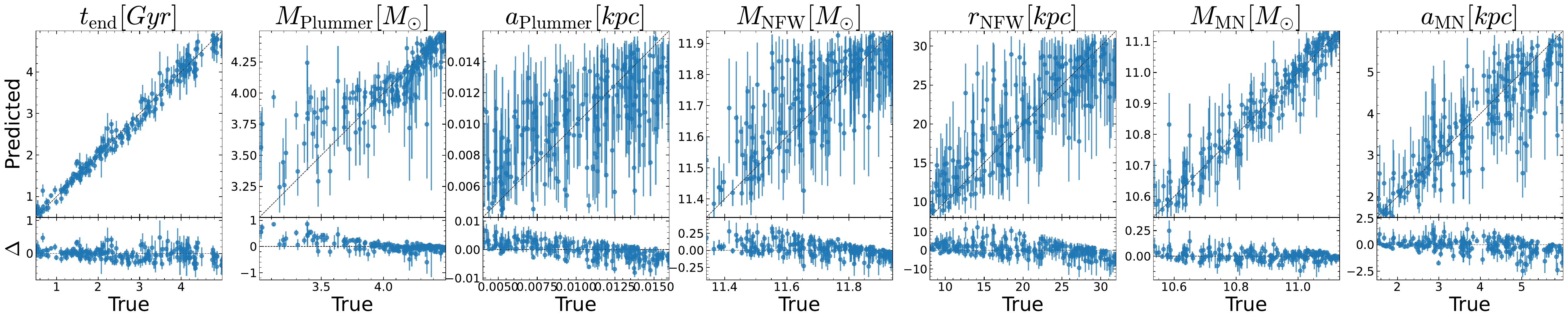}
    \vspace{0.5em} 
    \includegraphics[width=0.99\linewidth]{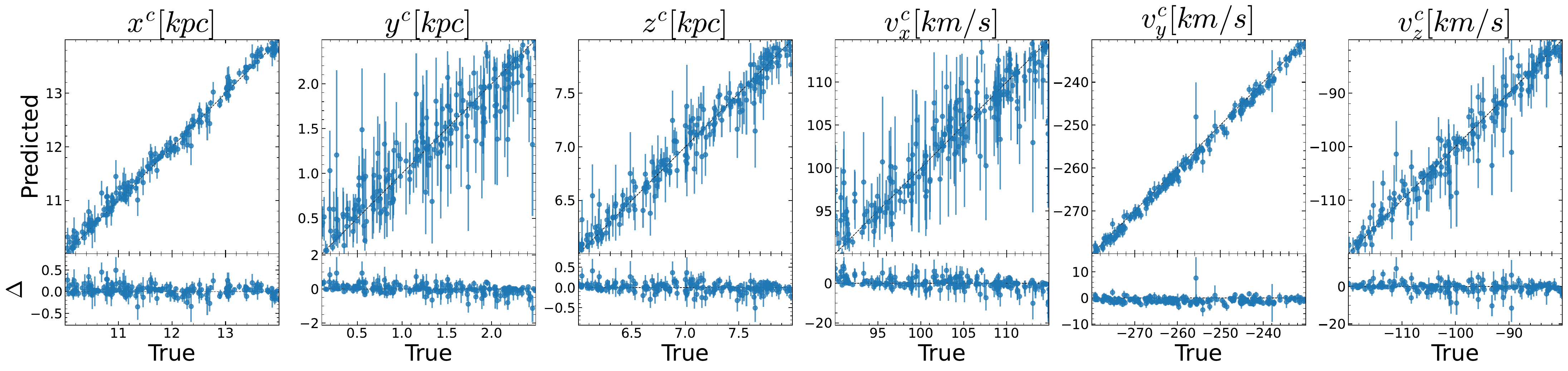}
    \caption{True-predicted plots for the test set. We subsampled the test set to 500 for visual clarity. The circles represent the median, while the error bars report the 16th to 84th percentiles. We also report the residuals $\Delta$, with error bars at the 16th to 84th percentiles.}
    \label{fig:predictions_combined}
\end{figure*}

\subsection{Architecture: \texttt{SetTransformer} }\label{subsection:architecture}
Given the permutation-invariant nature of our data—the set of $N$-body particles—we adopted the \texttt{SetTransformer} architecture introduced by \citep{lee_set_2019}. Our model extends this framework to infer the vector field $v_\phi$, conditioned on the flow matching integration time $t$, the model parameters $\theta_t$, and the particle observations $d$.

Each input was first embedded into a latent representation via multilayer perceptrons (MLPs). The integration time $t$ was further encoded through a sinusoidal time embedding following \citep{ho_denoising_2020}. The embedded particle features are processed by a stack of self-attention blocks (SABs) with residual connections, which capture interactions among particles. To incorporate temporal conditioning, the output of each SAB was modulated via feature-wise linear modulation (FiLM; \citealt{perez_film_2017}) using the embedded time representation. Before producing the vector field, a cross-attention block (CAB) was applied, where the query $Q$ corresponds to the embedded parameters $\theta_t$, and the keys $K$ and values $V$ are given by the SAB plus FiLM representation of the particle set. In this configuration, the model parameters act as queries that guide the attention toward relevant features within the particle embeddings. The cross-attention output was then modulated again through a FiLM layer conditioned on the embedded time. Finally, the predicted vector field $v_\phi$ was obtained via a last MLP layer. The architecture is described in Fig. \ref{fig:SetTransformer}.

\subsection{Training details}\label{subsection:training_details}
We generated \(2 \times 10^5\) stellar streams, each containing a fixed number of stars, N=1000\footnote.{We chose this number to have the same order of magnitude of stars as in GD-1 membership studies, such as the one presented in \citep{Ibata2024_DR3}.}  Of these, \(5 \times 10^4\) samples were used for validation and \(10^3\) for testing. In Appendix \ref{appendix: ablation_study} we report an ablation study used to verify that the number of simulations is enough to reach convergence of the training.
The training employed early stopping with a patience of 30 epochs, resulting in a total of 123 training epochs. We used a batch size of 500 and optimized the model with AdamW, starting from a learning rate of \(10^{-3}\) and applying a reduce-on-plateau schedule with a patience of five epochs. The final architecture, illustrated in Fig.~\ref{fig:SetTransformer}, was selected after extensive experimentation with different design choices and hyperparameters, including the number of neurons, the number of SABs and CAB blocks, activation functions, modulation schemes, and related configurations.
The training set generation took $\sim$ 7.5 GPU hours on an NVIDIA (Hopper) H200 (140 GB). The training and testing of the FMPE took $\sim$ 8 hours on an NVIDIA (Ampere) A100 (40 GB). The entire inference pipeline was carried out using the \texttt{sbi-sim}\footnote{\url{https://github.com/tum-pbs/sbi-sim/tree/dev}} package presented in \cite{holzschuh_flow_2024}. 

\section{Results}\label{sec:results}
We demonstrate the validity of the results of our inference by reporting a few metrics obtained on the test set. This set of $(d, \theta)$ was not used during training. For each of the test set couples, we sampled $10^3$ samples from the posterior distribution. In order to evaluate the accuracy and calibration over the whole test set, as we report in Sec \ref{subsec:calibration} and \ref{subsec:accuracy}, respectively, we predicted-true plots and percentile-percentile plots. In Sect. \ref{subsec:gd1} we report the posterior's samples obtained from inferences based on our mock observation of the GD-1 stream. For this test, we decided to sample $10^4$ times the posterior. Moreover, we performed posterior predictive checks to show that by forward passing these posterior samples, we obtain an observation that resembles the mock observation of GD-1.

\begin{figure*}
    \centering
    \includegraphics[width=0.99\linewidth, trim=5cm 2.5cm 8.8cm 2.5cm]{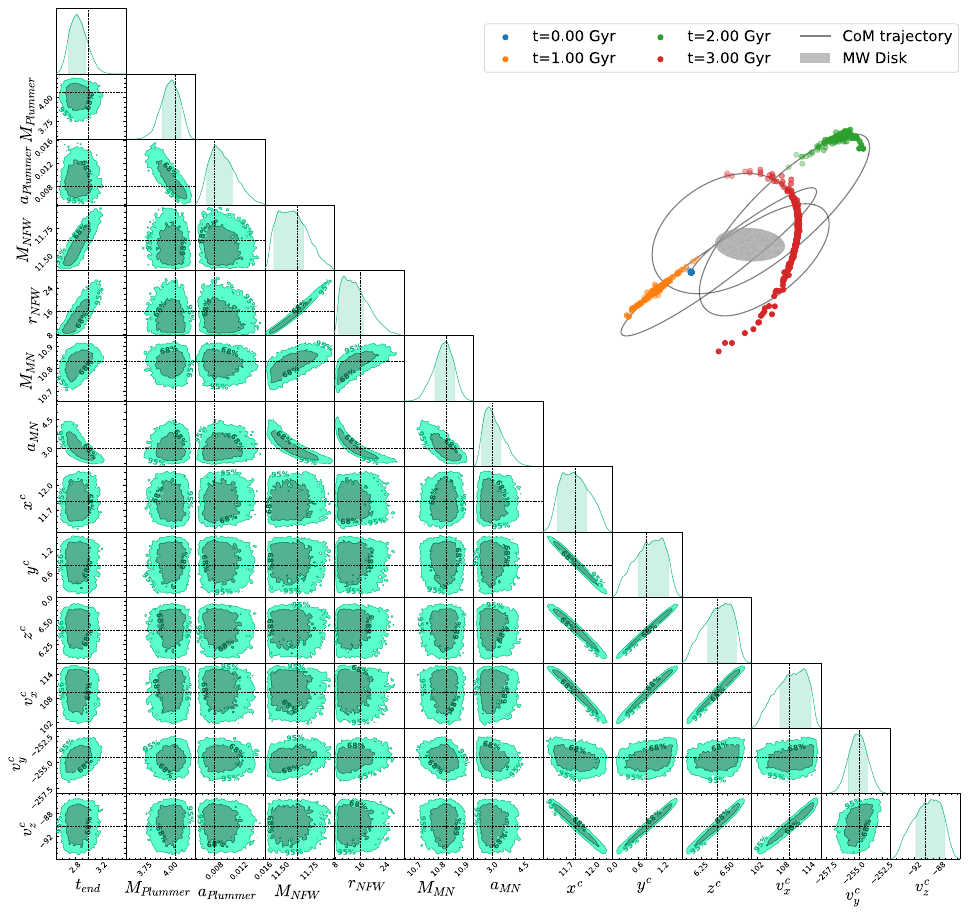}
    \caption{Lower left: Posterior samples from the fiducial GD-1 observation. The dashed line corresponds to the true value column in Table \ref{tab:priors}. Upper right: Snapshots of our fiducial GD-1 simulation. We reported an 8 kpc Milky Way disc for reference. }
    \label{fig:cornerplot}
\end{figure*}

\subsection{Model evaluation}

\subsubsection{Calibration}\label{subsec:calibration}
Percentile–percentile (P–P) plots are a diagnostic tool used to evaluate the calibration of posterior distributions in a probabilistic inference. The key concept is that a well-calibrated model should produce posteriors whose credible intervals contain the true parameter values with the correct empirical frequency. For the marginal distributions of the predicted posterior, the P–P plot visualizes this property by comparing the predicted percentiles under the inferred posterior against the empirical percentiles (the fraction of true parameters that fall below the corresponding posterior quantile). Ideally, if the posterior is perfectly calibrated, these two quantities coincide, and the curve follows the diagonal line. By inspecting these plots and their deviations from the diagonal, we can spot deviations from the true posterior distributions, such as over- or underconfidence and positive and negative biases. We report the marginal cover plots for individual parameters in Fig. \ref{fig:PP_plot} and the joint coverage plot in Fig.~\ref{fig:tarp}. One can appreciate that for the $\theta_{host}$, the P-P plots of marginal cover suggest a good calibration, while the two S-shape behavior in the Plummer parameters suggests a possible bias, which for the case of $M_{Plummer}$ is a positive bias as reflected by the over-estimation shown in Fig. \ref{fig:predictions_combined}. Moreover, a significant underconfidence (overconfidence) is shown for $x^c$ ($v^c_y$). 

However, marginal posterior coverages do not tell the whole story. We adopted the "tests
of accuracy with random point" (TARP) to study the behavior of the joint posterior distribution. As described in \cite{lemos_sampling-based_2023}, TARP provides a necessary and sufficient condition for the joint posterior coverage. The method constructs spherical credible regions around randomly chosen points in parameter space and measures the fraction of posterior samples contained within each region. By repeating this process over many random points, TARP estimates the empirical coverage as a function of the nominal credibility level. A perfectly calibrated joint posterior yields a one-to-one correspondence between nominal and empirical coverage, represented by a diagonal trend in the resulting TARP curve.
We report our Tarp in Fig. \ref{fig:tarp} to show we have good coverage on the joint distribution evaluated over the test set. TARP is sensitive to the geometry of the joint parameter space. Marginal P-P plots, on the other hand, assess one-dimensional projections obtained by integrating over all other parameters. These marginal checks can reveal parameter-specific biases or calibration issues that may not be immediately apparent in the joint coverage statistics, even when the overall joint posterior is well calibrated. Therefore, both diagnostics are complementary and indispensable for comprehensively assessing posterior coverage.

\subsubsection{Accuracy}\label{subsec:accuracy}
Having assessed that our posterior is well calibrated on almost all the marginals and the joint, we report the accuracy that we can expect by using our model in the form of true-predicted plots. In Fig. \ref{fig:predictions_combined}, the circles represent the median value, and the error bars represent the 16th to 84th percentile intervals. In the lower plots of the figure, we report the distribution of residuals to capture any possible residual trends. Our model seems to be extremely capable of capturing the posterior distribution for most of the parameters $\theta$, except for the $M_{Plummer}$ and $a_{Plummer}$, for which a mostly uniform trend seems to emerge when we inspect the distribution of the residuals. We interpret the mass bias as being due to the fact that under $\sim 10^3 M_\odot$, the feature that the progenitor can imprint of the stars is limited, and the stars are mostly dominated by the host potential. The Plummer scale $a_{Plummer}$ seems to play a less important role compared to other quantities, and the results of our analysis could be limited by the choice of a insufficiently broad prior.

\subsection{GD-1}\label{subsec:gd1}

In the following section we show the posterior distribution for our mock true observation of the GD-1 stream. This simulation is meant to allow for a direct comparison with the results of \citep{alvey_albatross_2024}, even though the set of parameters common between the two approaches is limited to ($t_{end}, M_{Plummer}, \vec{x}^c, \vec{v}^c)$. In Fig. \ref{fig:cornerplot}, we report the corner plot for the posterior. The results suggest that we can strongly constrain these parameters since the true value always lies inside the 16th to 84th percentile and that we can reproduce the results on \cite{alvey_albatross_2024}. We also captured the correlation between the host potential parameters, for example, ($M_{NFW}$,  $r_{NFW}$) and ($M_{MN}$, $a_{MN}$). These correlations are expected since different combinations lead to the same enclosed mass at a given radius. In Fig. \ref{fig:cornerplot}, one can also appreciate the correlations in the phase space $(x^c, v^c)$, a reflection of the spherical symmetry of the problem.
In agreement with the literature \citep{koposov_constraining_2010,nibauer_charting_2022,palau_constraining_2025}, these results show that stellar streams can probe the shape and amplitude of the gravitational potential.

\begin{figure}
    \centering
    \includegraphics[width=0.99\linewidth]{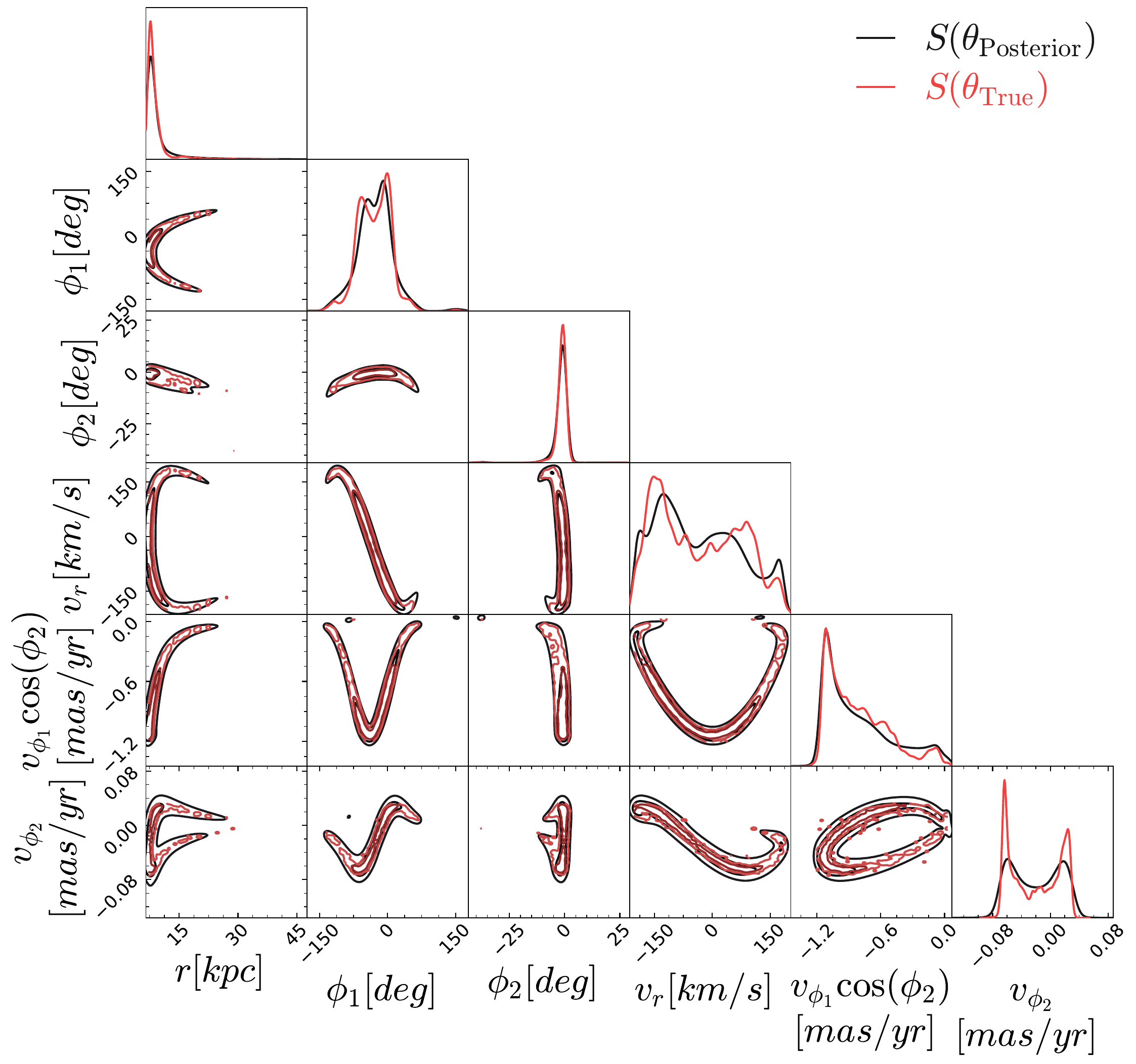}
    \caption{Posterior predictive check. We report in red our mock observation of the GD-1 stream obtained using the true values in Table \ref{tab:priors}. In black we report the joint results of forward modeling the 1000 samples in the posterior distribution. }
    \label{fig:PPC_corner}
\end{figure}

\begin{figure*}
    \centering
    \includegraphics[width=\linewidth]{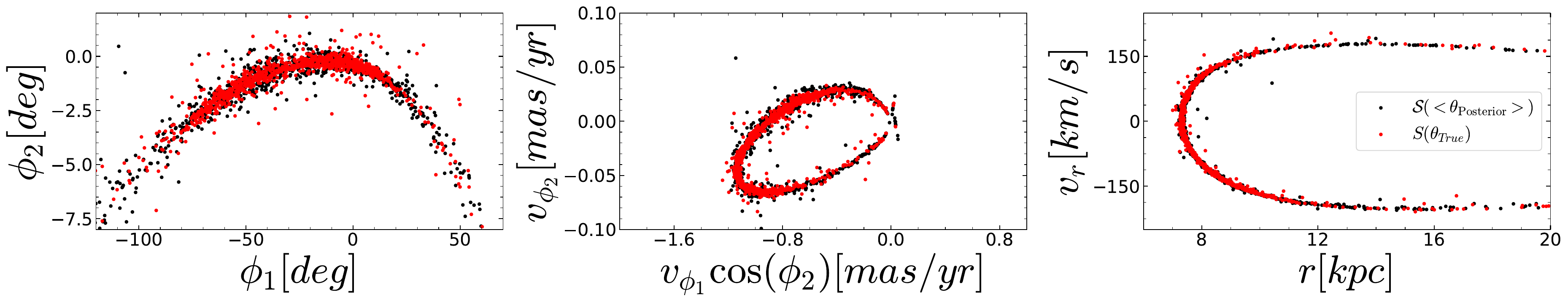}
    \caption{Posterior predictive check. We report in red our mock observation of the GD-1 stream just as in Fig. \ref{fig:PPC_corner}, and in black we show the forward model of the mean of the posterior samples, which is our estimate of the parameters $\theta$.}
    \label{fig:PPC_mean}
\end{figure*}

\paragraph{Posterior predictive check.}
As a complementary check, we implemented a forward pass on the posterior samples from Fig.~\ref{fig:cornerplot} to obtain $d \sim P(d \mid \theta_{\text{Posterior}})$. Our aim was to capture biases introduced by the inference, which would be reflected as a deviation of $d \sim P(d\mid\theta_{\text{Posterior}})$ from $d_{True}$. For each of the samples in Fig.~\ref{fig:cornerplot}, we generated a stellar stream with 1000 stars and simulated the tidal disruption as described in Sect. \ref{subsection:simulation_steps}. We then combined all the streams and checked if the distribution in observable space $d$ obtained by this procedure deviates from the stars in our mock observation. We report the results in Fig.~\ref{fig:PPC_corner}. In this plot, the "ground truth" is not a single point but a distribution of particles, so the black contours are given by displaying 1000 test set simulations, each with 1000 star particles. We note that all major stream features are captured nicely by our model. 
Moreover, in Fig. \ref{fig:PPC_mean} we report the observation obtained by forward modeling our estimate of the parameters and the posterior mean, which was obtained by the $N=1000$ samples in our posterior as $<\theta_{Posterior}> = 1/N\sum_{i=0}^N \theta_i$. Also in this test, we observed strong agreement with the mock observation.


\section{Discussion}\label{sec:Discussion}
\subsection{Comparison to classical GD-1 analyses}
The GD-1 stellar stream has been widely used to constrain the Milky Way gravitational potential while using a variety of dynamical techniques (see Sect.~\ref{sec:intro}), each making different approximations about stream formation and evolution. Orbit-fitting and backwards integration approaches have demonstrated that GD-1 provides strong constraints on large-scale properties of the Galactic potential, such as the circular velocity and overall flattening, but it typically approximates the stream as tracing a single orbit and therefore neglects its internal phase-space structure \citep[e.g.,][]{koposov_constraining_2010, malhan_constraining_2019,price-whelan_inferring_2014}. Action-angle based methods partially relax this assumption by modeling the stripping process in angle space, enabling improved recovery of stream morphology and yielding precise constraints on the shape of the gravitational force field at the stream location, although they often rely on analytic potential models and simplified prescriptions for mass loss \citep[e.g.,][]{bovy_dynamical_2014}. Particle-spray and stream-track techniques provide a compromise between physical fidelity and computational cost by approximating the escape of stars from the progenitor, but they do not capture the fully self-consistent dynamical evolution of the progenitor--stream system \citep[e.g.,][]{bowden_dipping_2015}.

In contrast, the framework presented here performs fully joint Bayesian inference on both progenitor properties and host-galaxy potential parameters using self-consistent $N$-body forward modeling combined with simulation-based inference. Rather than fitting a simplified representation of the stream, the method learns the mapping between physical parameters and the full phase-space distribution of stream stars. In controlled experiments, we have demonstrated accurate recovery of the full 13-dimensional parameter vector describing the Galactic potential and progenitor phase-space properties (see Fig.~\ref{fig:cornerplot}). We report a calibrated joint posterior (Fig.~\ref{fig:tarp}) and calibrated marginal posteriors for the host and for most progenitor parameters (see Fig.~\ref{fig:PP_plot} and Sect.~\ref{subsec:calibration}). This joint inference naturally captures degeneracies between host and progenitor parameters that are typically fixed or marginalized over in classical analyses. The principal trade-off is the upfront computational cost associated with generating training simulations and training the neural posterior estimator. However, once trained, the amortized nature of the method enables rapid posterior inference without additional simulations or likelihood evaluations. In addition, our GPU accelerated N-body code \textsc{\texttt{Odisseo}} \citep{Viterbo_2025} presents a very efficient way of performing stream simulations. In this sense, our approach complements classical techniques by prioritizing modeling fidelity, joint inference, and posterior calibration over analytic tractability.

\subsection{Model misspecification and robustness}
A fundamental assumption underlying SBI is that the observed data are generated by a process that lies within, or sufficiently close to, the family of simulators used for training. In the present work, the neural posterior estimator is trained exclusively on simulations generated using the \texttt{BovyMWPotential2014} functional form \citep{bovy_shape_2016}. If the true Milky Way potential deviates from this family, for example through halo triaxiality, time-dependent mass growth, or perturbations induced by massive satellites such as the Large Magellanic Cloud, the inference problem becomes formally misspecified or out-of-distribution. In such cases, the learned posterior is no longer guaranteed to be calibrated and may exhibit biased parameter estimates, underestimated uncertainties, or failures of posterior coverage, even if it performs well on in-distribution test simulations.

Recent work in the SBI literature has shown that model misspecification leads to characteristic statistical signatures that can be diagnosed using principled validation tools \citep[e.g.,][]{Schmitt2021,Zhou2025}. Simulation-based calibration, posterior predictive checks, and classifier-based two-sample tests can be used to assess whether the observed data are statistically compatible with the assumed simulator family. In our framework, posterior predictive simulations drawn from the inferred parameter distribution (e.g., Figs.~\ref{fig:PPC_corner} and~\ref{fig:PPC_mean}) can be directly compared to the observed stream morphology and phase-space structure, allowing systematic residuals or mismatches to be identified as indicators of forward-model incompleteness.

From an astrophysical perspective, unmodeled perturbations to the Galactic potential are expected to primarily affect localized or higher-order features of the stream, such as density variations, track offsets, or kinematic asymmetries, rather than uniformly shifting all global potential parameters. Consequently, model misspecification may manifest as broadened posteriors or altered parameter degeneracies rather than simple point-estimate biases. Importantly, the simulation-based nature of the present framework makes it straightforward to extend the training distribution to include more flexible or perturbed potential models (e.g., time-dependent halos, non-axisymmetric components, or explicit satellite perturbations) without requiring analytic likelihoods or changes to the inference machinery. Similarly, our N-body simulation based approach will be able to make use of prior constraints in the potential shape as informed by cosmological simulations \citep[e.g.,][]{Obreja2022} and will not be limited to stellar streams alone. It can similarly make use of any halo substructure as found by state-of-the-art structure finders \citep{Oliver2024,Oliver2024a}. Furthermore, our approach will not be limited to the Milky Way alone but might as well be applied to the large sample of external galaxies hosting stellar streams \citep[e.g.,][]{Miro2025}. The present work therefore provides a controlled baseline within which robustness to model misspecification can be systematically explored in future studies.

\subsection{Pathway toward application to real GD-1 data}
While our current work focuses on an idealized simulation setting, the framework is designed with application to real GD-1 observations in mind. Extending the analysis to realistic survey data primarily requires augmenting the forward model rather than modifying the inference procedure. In practice, this involves sampling stellar magnitudes from the observed luminosity function of GD-1, applying survey-specific selection functions, and resampling phase-space coordinates with magnitude-dependent uncertainties appropriate for Gaia and complementary spectroscopic surveys. Background contamination can be incorporated by forward modeling a mixture of stream and field stars, allowing the neural posterior estimator to marginalize over membership uncertainty. Conceptually, these extensions correspond to replacing the idealized observation block in the SBI pipeline shown in Fig.~\ref{fig:flow_matching_flowchart} with a survey-aware observational model while leaving the remaining components unchanged, as also demonstrated in recent SBI-based dynamical analyses \citep[e.g.,][]{nibauer_charting_2022,alvey_albatross_2024}.

Specifically, the choice to fix the number of stars per simulation to $N=1000$ was motivated by computational considerations and by the desire to match the order of magnitude of the current GD-1 membership samples. This choice does not represent a fundamental limitation of the method. The \texttt{SetTransformer}former architecture employed here is permutation invariant and can, in principle, handle variable-size inputs. In future applications, robustness to varying sample sizes can be achieved by drawing the number of stars per simulation from a distribution informed by the survey selection function or by randomly subsampling from larger simulated streams during training. These strategies enable the model to learn invariance to completeness and membership size, which will be essential when analyzing real observational data.


\section{Conclusion}\label{sec:Conclusion}
With this work, we have presented the first application of SBI to the GD-1 stream to jointly constrain parameters of the progenitor and the host potential. With the \texttt{\textsc{Odisseo}} simulator, we ran a large set of mock GD-1 stream analogs to reproduce the tidal stripping of a globular cluster, which we used to train our neural density estimator. We adopted a \texttt{SetTransformer} architecture to automatically extract summary statistics from the observations $x$, enabling a robust inference implemented via a flow matching strategy. Our results are summarized below.

\begin{itemize}
    \item We have produced a large set of publicly available mock GD-1 streams using the \texttt{\textsc{Odisseo}} simulator. 
    \item We have extended the work of \citep{alvey_albatross_2024} to also study the gravitational potential of the host galaxy, achieving an amortized inference that does not require sequential training.  
    \item We have shown the validity of our model with extensive testing in the form of P-P plots for coverage, true-predicted plots for accuracy, and posterior predictive checks for self-consistency.
    \item We recovered the fiducial parameters over our fiducial GD-1 observation in Fig \ref{fig:cornerplot} for this controlled experiment. However, performance under more realistic survey conditions remains to be tested. We captured strong correlations between mass and scale length parameters $(M_{NFW}-r_{NFW}; M_{MN}-a_{MN}; r_{NFW}-a_{MN})$ and, surprisingly, a weak positive correlation between the masses $(M_{NFW}-M_{MN})$. We also captured a strong correlation between the present-day phase space of the progenitor of GD-1 $(\vec{x}^c, \vec{v}^c)$, a result of the axisymmetry of the problem. A notable exception is $v_y^c$, for which our model is also not well calibrated, as can be observed in Fig. \ref{fig:PP_plot}.
    \item We leveraged the freedom given by the flow matching technique to adopt a robust transformer architecture capable of handling this high dimensional inference task. 
\end{itemize}

Our aim was to quantify, in a controlled simulation setting, the information that a single cold stellar stream (GD-1 analogs) carries about the progenitor properties and the host galaxy potential. We did so by combining self-consistent N-body forward modeling (\textsc{\texttt{Odisseo}}) with amortized likelihood-free inference (flow matching) and by validating the calibration and accuracy on held-out synthetic data.

In future work, we will extend this method with a more robust, realistic, and survey-dependent handling of the observational errors, which we have simplified in this work; magnitude-dependent selection functions; and background contamination, which were both ignored. We also intend to extend the pipeline to leverage multistream observation, inspired by the results obtained in \citep{bovy_shape_2016}. 
Lastly, since \texttt{\textsc{Odisseo}} is capable of calculating the gradient of the simulation with respect to the input parameters, we aim to incorporate this as additional information to guide the inference pipeline, as shown in \citep{holzschuh_flow_2024}.

\section*{Data availability}\label{sec:appendix_code}

We publicly release our code to reproduce all the figures via GitHub: \url{https://github.com/vepe99/sbi-sim/tree/odisseo_branch}. 

\begin{acknowledgements}
      This project was made possible by funding from the Carl-Zeiss-Stiftung.
      This work was supported by the Deutsche Forschungsgemeinschaft (DFG, German Research Foundation) under Germany’s Excellence Strategy EXC 2181/1 - 390900948 (the Heidelberg STRUCTURES Excellence Cluster). We acknowledge the usage of the AI-clusters Tom and Jerry funded by the Field of Focus 2 of Heidelberg University. 
      We thank William H. Oliver, Lorenzo Branca, Leonard Storcks and Anna Lena Schaible  for insightful discussions and stimulating conversations that helped shape the ideas presented in this work.
      We are grateful to Paola Ziero for the design of the project logo and her creative support. 
\end{acknowledgements}

\bibliographystyle{aa}
\bibliography{bibtex/bib}

\appendix
\onecolumn
\section{Ablation study}
\label{appendix: ablation_study}
\begin{figure*}[h]
    \centering
    \includegraphics[width=\linewidth]{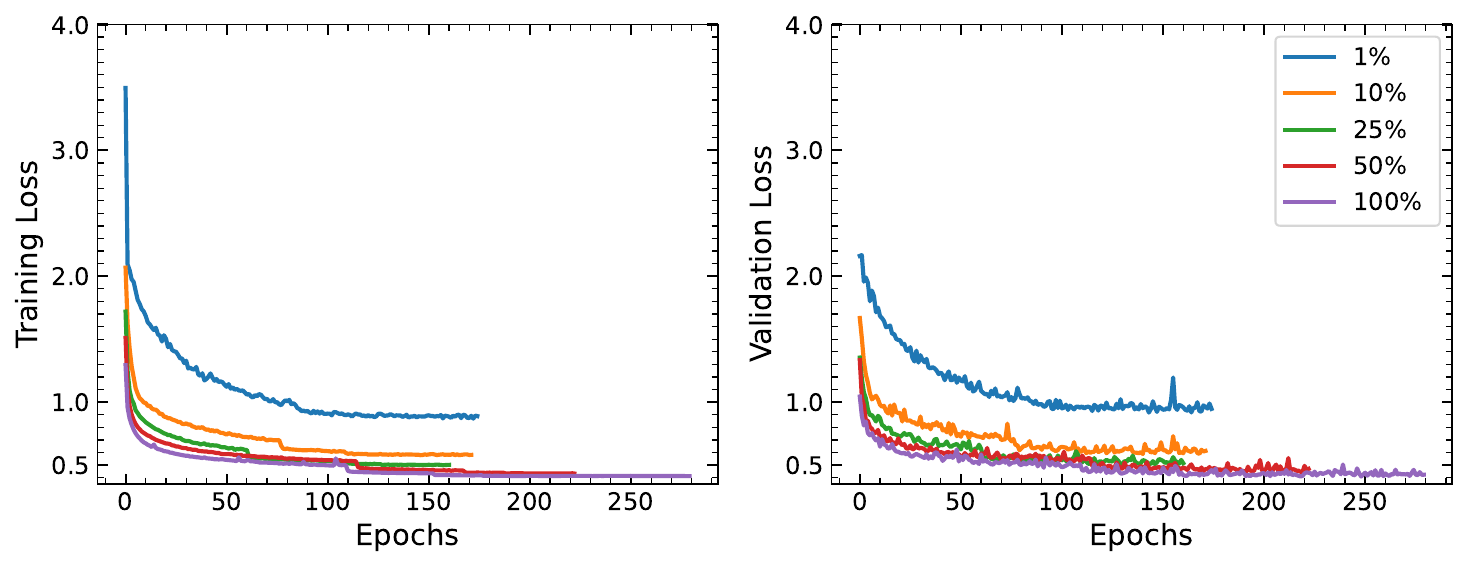}
    \caption{Ablation study showing training loss (left) and validation loss (right) as a function of training epochs for different simulation budgets (1\%, 10\%, 25\%, 50\%, and 100\% of the full training set). All models converge within ~100-150 epochs, with larger simulation budgets achieving better final performance.}
    \label{fig:ablation}
\end{figure*}

Figure \ref{fig:ablation} presents an ablation study examining how training and validation loss evolve as a function of simulation budget, defined as the fraction of the available training set used for model training.
The ablation study demonstrates clear convergence behavior across all simulation budgets. Both training and validation losses decrease monotonically and stabilize after approximately 100-150 epochs, indicating successful model optimization. As expected, larger simulation budgets (50\% and 100\%) achieve lower final validation losses ($\sim$0.4-0.5), confirming that increased training data improves generalization performance. While comprehensive evaluation of an SBI pipeline requires more sophisticated metrics (see Sects. \ref{subsec:calibration} and \ref{subsec:accuracy}), this ablation study justifies our choice of training set size.

\end{document}